\documentclass[fleqn,12pt,twoside]{article}
\usepackage{espcrc1}

\usepackage{graphicx}

\newcommand{\AmS}{{\protect\the\textfont2
  A\kern-.1667em\lower.5ex\hbox{M}\kern-.125emS}}

\title{Jet Quenching with Detailed Balance in a Thermal QCD Medium
\thanks{This work was supported by the
NSFC under projects Nos. 19928511 and 10135030, and by the U.S.
Department of Energy under Contract No. DE-AC03-76SF00098.}}

\author{Enke Wang\address[IOPP]{Institute of Particle Physics,
        Huazhong Normal University,
        Wuhan 430079, China}
        and Xin-Nian Wang
        \address{Nuclear Science Division, Lawrence Berkeley Laboratory,
         Berkeley, California 94720}\addressmark[IOPP]}

\begin{document}

\maketitle

\begin{abstract}
The effect of detailed balance in induced gluon radiation for an energetic
parton propagating inside a quark-gluon plasma is discussed. The 
stimulated thermal gluon emission reduces while absorption increases 
the parton's energy. The net contribution results in a reduction of 
the effective parton energy loss. The
relative reduction of the parton energy loss is found to be
important for intermediate parton energies and negligible for very
high energies. This will increase the energy dependence of the 
parton energy loss and will affect the shape of suppression of 
moderately high $p_T$ hadron spectra.
\end{abstract}

\section{INTRODUCTION}

Jet quenching in high-energy heavy-ion collisions suppresses
large transverse momentum hadrons as compared to $pp$ collisions
at the same energy \cite{WG92,GW94}. It provides a direct 
probe of the dense matter produced
in high-energy heavy-ion collisions. Recent theoretical studies
\cite{BDPMS,Zakharov,GLV,GuoW,Wied} of jet quenching have
concentrated on gluon radiation induced by multiple scattering in
medium. Because of the presence of thermal gluons in the hot
medium the stimulated gluon emission and absorption by the
propagating parton should also be taken into account consistently. Such
a detailed balance is crucial for parton thermalization and should
also be important for calculating the energy loss of an energetic
parton in quark-gluon plasma (QGP)\cite{WW01}. We will show that
the partial cancellation between stimulated emission and thermal
absorption results in a net reduction of parton energy loss
induced by multiple rescattering. Though
such a reduction is negligible for asymptotically large parton
energy, it is still important for intermediate values of energy.

\section{FINAL-STATE AND RESCATTERING-INDUCED ABSORPTION}

In an axial gauge
and in the leading-log approximation, the final-state radiation
amplitude off a quark can be factorized from the hard scattering.
Taking into account of both stimulated emission and thermal
absorption in a thermal medium with finite temperature $T$, 
the probability of gluon radiation with energy $\omega$ can be
written as
 \begin{equation}
 \label{prob0}
   {{dP^{(0)}}\over d\omega}={{\alpha_s C_F}\over 2\pi}
     \int {{dz}\over z} \int
     {{d{\bf k}_{\perp}^2}\over {\bf k}_{\perp}^2}
     \Big[ N_g(zE)\delta(\omega+zE)+
     \left(1+N_g(zE)\right)\delta(\omega-zE)\theta(1-z)\Big]
     P({\omega\over E})\, ,
 \end{equation}
where $N_g(|{\bf k}|)=1/[\exp(|{\bf k}|/T)-1]$ is the thermal
gluon distribution, $C_F$ is the Casimir of the quark jet in the
fundamental representation and the splitting function
$P_{gq}(z)\equiv P(z)/z=[1+(1-z)^2]/z$ for $q\rightarrow gq$. In
Eq.(\ref{prob0}) the first term is from thermal absorption and the
second term from gluon emission with the Bose-Einstein enhancement
factor.

To define the effective parton energy loss, we only consider gluon
radiation outside a cone with $|{\bf k}_{\perp}|>\mu$. Here $\mu$
should be a nonperturbative scale for parton fragmentation in the
vacuum. In a hot QGP medium, it can be replaced by the gluon's
Debye screening mass. Assuming the scale of the hard scattering as
$Q^2=4E^2$, one has the kinematic limits of the gluon's transverse
momentum, $\mu^2 \leq {\bf k}_{\perp max}^2 \leq
4|\omega|(E-\omega) \, .$

Subtracting the gluon radiation spectrum in the vacuum, one then
obtains the energy loss due to final-state thermal absorption and
stimulated emission. For asymptotically large parton energy, $E\gg
T$, we have
 \begin{eqnarray}
 \label{eloss0}
   {\Delta E^{(0)}_{abs}\over E} &=& \int d\omega \,
   {\omega\over E}
   \left({{dP^{(0)}}\over d\omega}-
     {{dP^{(0)}}\over d\omega}\Big|_{T=0}\right)
 \nonumber\\
&\approx& -
   {{\pi\alpha_s C_F}\over 3}
   {T^2\over E^2}\left[
     \ln{4ET\over \mu^2}+2-\gamma_{\rm E}
     +{{6\zeta^\prime(2)}\over \pi^2}\right],
\end{eqnarray}
where $\gamma_{\rm E}\approx 0.5772$ and $\zeta^\prime(2)\approx
-0.9376$.

During the propagation of the hard parton after its
production, it will suffer multiple scattering with targets in the
medium. Recently many theoretical studies
\cite{BDPMS,Zakharov,GLV,GuoW,Wied} have investigated the 
rescattering-induced radiative energy loss. Here we will focus on the
stimulated emission and thermal absorption associated with
rescattering in a hot QGP medium.

Assume a hard parton is produced at ${\bf y}_0=(y_0, {\bf
y}_{0\perp})$ inside the medium with $y_0$ being the longitudinal
coordinate. The interaction between the jet and target partons is
modeled  by a static color-screened Yukawa potential which is
initially proposed by Gyulassy-Wang (GW) \cite{GW94},
 \begin{equation}
 \label{potential}
   V_n
    =2\pi\delta(q^0)v({\bf q_n})e^{-i{\bf q}_n\cdot {\bf y}_n}
       T_{a_n}(j)T_{a_n}(n)\, , \qquad
  v({\bf q}_n)={{4\pi\alpha_s}\over {{\bf q}^2_n+\mu^2}}\, .
 \end{equation}
Here ${\bf q}_n$ is the momentum transfer from a target parton $n$
at ${\bf y}_n=(y_n, {\bf y}_{\perp n})$,
 $T_{a_n}(j)$ and $T_{a_n}(n)$ are the color matrices for the jet and
target parton.

Following the framework of opacity expansion developed by Gyulassy,
L\'evai and Vitev (GLV)\cite{GLV} and Wiedemann \cite{Wied},
we will only consider the contributions to the first order in the
opacity expansion. It was shown by GLV that the higher order
corrections contribute little to the radiative energy loss.
Including stimulated emission and thermal absorption we can
express the radiation probability at the first order in opacity as
 \begin{eqnarray}
 \label{prob1}
   {{dP^{(1)}}\over d\omega}&=&{{C_2}\over {8\pi d_A d_R}}
     \int {{dz}\over z}  \int
      {{d^2{\bf k}_{\perp}}\over {(2\pi)^2}}
     \int {{d^2{\bf q}_{\perp}}\over {(2\pi)^2}}
      v^2({\bf q}_{\perp}){N\over A_{\perp}}
      \left\langle Tr\left[|R^{(S)}|^2
     {+}2 Re\left(R^{(0)\dagger}
     R^{(D)}\right)\right]\right\rangle
 \nonumber\\
     &&\times P({\omega\over E})\Big[\left(1+N_g(zE)\right)\delta(\omega-zE)\theta(1-z)
     +N_g(zE)\delta(\omega+zE)\Big]\, ,
 \end{eqnarray}
where $C_2$ and $C_A$ are Casimirs of the target parton in the
fundamental representations in $d_R$ and $d_A$ dimension.
Averaging over the longitudinal target profile is defined as
$\left\langle \cdots\right\rangle=\int dy \rho(y)\cdots$ with
$\rho(y)=2 \exp(-2y/L)/L$. We denote $R^{(0)}$, $R^{(S)}$ and
$R^{(D)}$ as the radiation amplitude associated with zero, single
and double Born rescattering with parton targets in the QGP,
respectively. They can be obtained as\cite{GLV,WW01}
 \begin{eqnarray}
 \label{amps}
   R^{(0)}&=&2ig T_c{{{\bf k}_{\perp}\cdot{\bf \epsilon}_{\perp}}
      \over {\bf k}^2_{\perp}}\, ,\\
   R^{(S)}&=&2ig \Bigl( {\bf H}T_a T_c
     +{\bf B}_1 e^{i\omega_0 y_{10}}[T_c, T_a]
     +{\bf C}_1 e^{i(\omega_0-\omega_1) y_{10}}[T_c, T_a]\Bigr)
     \cdot \epsilon_{\perp}\, ,\\
   R^{(D)}&=&2ig T_c e^{i\omega_0y_{10}}
     \Bigl(-{{C_F+C_A}\over 2}{\bf H} e^{-i\omega_0y_{10}}
     +{C_A\over 2}{\bf B}_1
     +{C_A\over 2}{\bf C}_1 e^{-i\omega_1y_{10}}\Bigr)
     \cdot \epsilon_{\perp}\, .
\end{eqnarray}
Here $\epsilon_{\perp}$ is the transverse polarization of the
radiative gluon, $y_{10}=y_1-y_0$,
 \begin{eqnarray}
 \label{omega}
   \omega_0&=&{{{\bf k}_{\perp}^2}\over {2\omega}}\, ,
   \quad
   \omega_1={{({\bf k}_{\perp}-{\bf q}_{\perp})^2}\over
      {2\omega}}\, ,
 \\
   {\bf H}&=&{{\bf k}_{\perp}\over {{\bf k}_{\perp}^2}}\, ,
   \quad
   {\bf C}_1={{{\bf k}_{\perp}-{\bf q}_{\perp}}\over
      ({{\bf k}_{\perp}-{\bf q}_{\perp}})^2}\, ,
   \quad
   {\bf B}_1={\bf H}-{\bf C}_1\, .
 \end{eqnarray}

The zero-temperature part of the parton energy loss corresponds to
the radiation induced by rescattering which should be the same as
obtained in the previous studies \cite{GLV}. We denote this part
as $\Delta E^{(1)}_{rad}$. The remainder or temperature-dependent
part of energy loss induced by rescattering at the first order in
opacity is
\begin{eqnarray}
 \label{eloss1}
   \Delta E^{(1)}_{abs}&=&\int d\omega \,\omega
     \left({{dP^{(1)}}\over d\omega} -{{dP^{(1)}}\over d\omega}
       \Big|_{T=0}\right)
   \nonumber\\
   &=&{{\alpha_s C_F}\over \pi}{L\over \lambda_g}E
   \int dz \int {d{\bf k}_{\perp}^2\over {\bf k}_{\perp}^2}
   \int d^2{\bf q}_{\perp}
    |{\bar v}({\bf q}_{\perp})|^2
      {{{\bf k}_{\perp}\cdot {\bf q}_{\perp}}\over
      {\left({\bf k}_{\perp} - {\bf q}_{\perp}\right)^2}} N_g(zE)
 \nonumber\\
     &&\times\Bigl[-P(-z)\left\langle
     Re(1-e^{i\omega_1 y_{10}})\right\rangle
     +P(z)\left\langle
     Re(1-e^{i\omega_1 y_{10}})\right\rangle\theta(1-z) \Bigr]\, ,
\end{eqnarray}
where $\lambda_g=C_F\lambda/C_A$ is the mean-free-path of the
gluon, the factor $1-\exp(i\omega_1 y_{10})$ reflects the
destructive interference arising from the non-Abelian 
Landau-Pomeranchuck-Migdal (LPM) \cite{LPM} effect, 
and $|{\bar v}({\bf q}_{\perp})|^2$ is the
normalized distribution of momentum transfer from the scattering
centers,
 \begin{equation}
 \label{vbar}
  |{\bar v}({\bf q}_{\perp})|^2 =
  {1\over \pi}{\mu^2_{eff}\over ({\bf q}_{\perp}^2+\mu^2)^2}\, ,
 \quad
   {1\over \mu^2_{eff}} = {1\over \mu^2}-{1\over
   q_{\perp max}^2+\mu^2}\, ,\quad
   q_{\perp max}^2\approx 3E\mu.
 \end{equation}

 In the limit of $EL\gg 1$ and $E\gg \mu$, One can
then get the approximate asymptotic behavior of the energy loss,
 \begin{eqnarray}
 \label{elossem3}
   {\Delta E_{rad}^{(1)}\over E} &\approx &
   {{\alpha_s C_F \mu^2 L^2}\over 4\lambda_gE}
   \left[\ln{2E\over \mu^2L} -0.048\right]\, ,
 \\
 \label{elossab3}
   {\Delta E_{abs}^{(1)}\over E} &\approx &-
   {{\pi\alpha_s C_F}\over 3} {{LT^2}\over {\lambda_g E^2}}
   \left[
   \ln{{\mu^2L}\over T} -1+\gamma_{\rm E}-{{6\zeta^\prime(2)}\over\pi^2}
\right] .
 \end{eqnarray}
Similarly to the final-state absorption, the minus sign in
Eq.~(\ref{elossab3}) indicates that the contribution from thermal
absorption associated with rescattering is larger than that of
stimulated emission, resulting in a net energy gain.


\begin{figure}[htb]
\label{fig1}
\begin{minipage}[t]{75mm}
\includegraphics[scale=0.4]{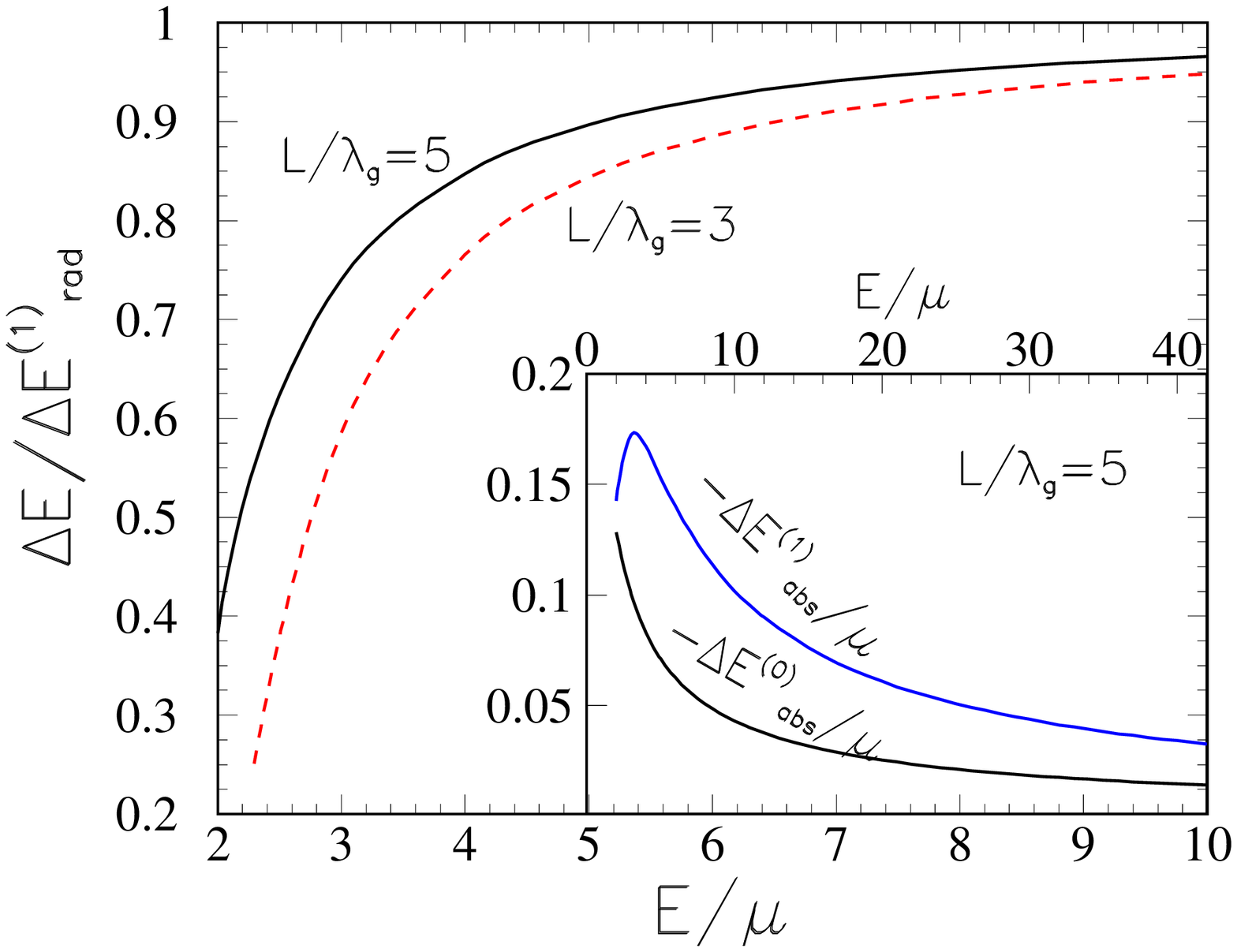}
\end{minipage}
\hspace{\fill}
\begin{minipage}[b]{70mm}
Figure 1. The ratio of effective parton energy loss with ($\Delta
E=\Delta E^{(0)}_{abs}+\Delta E^{(1)}_{abs} +\Delta
E^{(1)}_{rad}$) and without ($\Delta E^{(1)}_{rad}$) absorption as
a function of $E/\mu$. Inserted box: energy gain via absorption
with ($-\Delta E^{(1)}_{abs}$) and without ($-\Delta
E^{(0)}_{abs}$) rescattering.
\end{minipage}
\end{figure}

\section{NUMERICAL RESULTS AND CONCLUSION}

The significance of the thermal absorption relative to the induced
radiation can be studied by evaluating Eqs.~(\ref{eloss0}) and
(\ref{eloss1}) numerically. From the perturbative QCD at finite
temperature the Debye screening mass can be taken as
$\mu^2=4\pi\alpha_s T^2$ \cite{HTL}. In the GW model the
mean-free-path for a gluon $\lambda_g$ is \cite{GW94}
 \begin{equation}
 \label{lambda}
    \lambda_g^{-1}= \langle\sigma_{qg}\rho_q\rangle
+\langle\sigma_{gg}\rho_g\rangle
    \approx {2\pi\alpha^2_s\over \mu^2} 9\times 7\zeta(3)
    {T^3\over \pi^2}\, ,
 \end{equation}
where $\zeta(3)\approx 1.202$. With fixed values of $L/\lambda_g$
and $\alpha_s$, $\Delta E/\mu$ is a function of $E/\mu$ only.

The ratios of the calculated radiative energy loss with and
without stimulated emission and thermal absorption as functions of
$E/\mu$ are illustrated in Fig.~1 for $L/\lambda_g=3$,5 and
$\alpha_s=0.3$. The thermal absorption reduces the effective
parton energy loss by about 30-10\% for intermediate values of
parton energy. The energy dependence of the effective parton
energy loss becomes stronger. This will affect the shape of suppression
of $p_T$ hadrons spectrum in the intermediate energy region. For
partons with very high energy the effect of the gluon absorption
is small and can be neglected.

\end{document}